
%
%
%

\font\bbigten=cmr10 scaled\magstep2

\magnification=1200
\baselineskip=8truemm plus2truemm minus1truemm
\vsize=22.1truecm
\hsize=15truecm
\hoffset=0truecm
%
\def\qad@rato#1#2{{\vcenter{\vbox{
    \hrule height#2pt
    \hbox{\vrule width#2pt height#1pt \kern#1pt \vrule width#2pt}
    \hrule height#2pt} }}}
\def\qadratello{\mathchoice
 \qad@rato{5}{.5}\qad@rato{5}{.5}
 \qad@rato{3.5}{.35}\qad@rato{2.5}{.25} }
%
%
%
\null
\bigskip
\centerline{\bbigten Black Holes and the Adiabatic Phase}
\medskip
\centerline{Roberto Casadio and Giovanni Venturi}
\smallskip
\centerline{\it Department of Physics, University of Bologna, and}
\centerline{\it Istituto Nazionale di Fisica Nucleare, Sezione di
            Bologna, Italy}
\bigskip
\centerline{Abstract}
{\narrower{
An open system consisting of a scalar field bound to a
Kerr black
hole whose mass ($M$) and specific angular momentum ($a$) are slowly
(adiabatically) perturbed is considered.
The adiabatically induced phase and the conditions for the
validity of the adiabatic approximation are obtained.
The effect of closed cycles in parameter space ($a$, $M$ plane)
on the energy levels of both stable and unstable scalar field
bound states, together with other quantities of interest, is
illustrated.
Lastly it is noted that the black hole wavefunction
will acquire
an equal and opposite phase to that of matter thus leading to
a change of its effective action (entropy).
\smallskip}}
\par
\vfill
\eject
\centerline{\bf I. Introduction}
\medskip
As a consequence of the invariance of the action under arbitrary
space--time transformations, time does not appear in the Hamiltonian
form of gravity and such a feature is maintained in a quantum
formulation [1].
It has been shown, however, that the introduction of matter
and the semiclassical approximation for gravity allows one to
introduce the concept of time:
time parametrizes how matter follows gravity.
In particular it has been noted that the semiclassical wave function
for gravity provides a parametrization for the evolution of matter
in which the latter follows the former adiabatically [2,3].
\par
That one may consider an adiabatic approximation to the
matter--gravity system is a consequence of the mass scale of matter
being much less than the Planck mass.
Indeed for the matter--gravity system one may consider the generalized
coordinates associated with the former as being the ``fast'' variables
and those associated with the latter as being the ``slow'' variables.
\par
The adiabatic approximation has been studied in a number of different
contexts.
In the Born--Oppenheimer studies of molecules one has a Hamiltonian
which contains ``slow'' and ``fast'' degrees of freedom and in
particular the internuclear distance which is ``frozen'' in the
adiabatic process is regarded as a dynamical variable [4].
Alternatively one may just consider a Hamiltonian (for the ``fast''
variable) depending on a slowly varying external parameter
(``slow'' variable) in which case
the wave function acquires an additional phase and associated
``gauge'' connection with respect to the conventional dynamical one
[5].
In both approaches such an additional phase, which is a
function of the slowly varying heavy degrees of freedom, arises and for
closed loops cannot be eliminated in the presence of non--trivial
mappings of the ``gauge'' connection
into the heavy parameter space.
One will then get generalized Aharonov--Bohm effects.
This has been previously discussed for the case of the matter--gravity
system and in particular for the case of the three-geometry,
corresponding to heavy degrees of freedom, considered as a dynamical
variable [6].
\par
The purpose of this paper is to further study the adiabatically
induced phase factor arising from slowly varying three-geometries
where a time variable has been introduced as mentioned previously.
In particular we consider the case of a massive scalar field
bound to a rotating black hole whose mass ($M$) and specific
angular momentum ($a=J/M$) are varied.
That such variations can be achieved {\sl in principle}
has been previously illustrated:
for example, in ref. [7] a Kerr black hole has been used
as a Carnot engine.
We shall then imagine that through some analogous mechanism
the three--geometry of a rotating black hole is varied, the
variation being small and maintaining axial symmetry,
and see how such changes affect the states of a scalar field of mass
(Compton wavelength) much smaller (bigger) than the black
hole mass (size).
This is of interest since the performance of such a closed cycle
and the adiabatically induced phase will lead to a shift of the
black hole bound energy levels of the scalar field.
Further we note that the wave function for the complete system
(black hole plus scalar field) will consist of the product
of the black hole (slow degrees of freedom) and matter
(fast degrees of freedom) wave functions and it is known that
for closed cycles the two wave functions will acquire equal and
opposite phases [3,4,6].
Thus once we know the phase acquired by the matter wave function
we shall know that acquired by the black hole even if we do not
know its wave function (except in the semiclassical limit
where it is related to an exponential of the classical
gravitational action).
\par
In the next Section we shall obtain an effective action and
Hamiltonian for the scalar field moving in the Kerr metric
and subsequently obtain an expression for the adiabatic phase
together with an indication of the effect of fluctuations.
Lastly in Section III we examine and estimate the effects of
the phase and summarize our results.
In this paper we choose units such that $c=G=1$ and the metric
signature is ($-,+,+,+$).
\vfill
\eject
\centerline{\bf II. Scalar field and time dependent Kerr geometry}
\medskip
We shall consider the case of a massive scalar field bound to
a rotating black hole associated with a metric of the standard
Boyer--Lindquist [8] form:
$$
ds^2=-{\rho^2\,\Delta\over B}\,dt^2
     +{B\,\sin^2\theta\over\rho^2}\,\biggl(
      d\phi-{2\,a\,M\,r\over B}\,dt\biggr)^2
     +{\rho^2\over\Delta^2}\,dr^2
     +\rho^2\, d\theta^2
\ ,
\eqno(2.1)
$$
with:
$$
B=(r^2+a^2)^2-a^2\,\Delta\,\sin^2\theta
\eqno(2.2)
$$
$$
\rho^2=r^2+a^2\,\cos^2\theta
\eqno(2.3)
$$
and:
$$
\Delta=r^2-2\,M\,r+a^2\equiv(r-r_-)\,(r-r_+)
\eqno(2.4)
$$
where $r_{\pm}=M\pm\sqrt{M^2-a^2}$ are the radial coordinates
of the two horizons, the outer one ($r_+$) being the event
horizon for the Boyer--Lindquist observer we have chosen.
In what follows we shall always assume that $a<M$, thus
avoiding the extremal case ($a=M$).
Further as we have mentioned in the Introduction we shall
allow time dependent perturbations of the mass ($\delta M$)
and the Kerr specific angular momentum ($\delta a$).
In particular, on defining for brevity
$(\gamma^0,\gamma^1)=(\delta M,\delta a)$
we get:
$$
M(t)=M+\gamma^0
\eqno(2.5)
$$
$$
a(t)=a+\gamma^1
\ .
\eqno(2.6)
$$
This choice of coordinates is due to our desire to analyse
bound states of microscopic black holes which can most
easily be studied from the point of view of ``distant observers''
such as those of the Boyer--Lindquist type.
Further it is for such a choice of coordinates that the equation
of motion for the scalar field can be related, in an opportune
limit, to the same equation as governs an electron in the hydrogen
atom [10].
Thus, while the physics involved does not depend on the coordinates,
other choices would not be suitable for our analysis and would
lead to unnecessary complications.
\par
Let us now discuss the matter action.
It will be given by (disregarding surface terms):
$$\eqalignno{
S^{^M}&={1\over2}\int dt d^3x\, \sqrt{-g}\,
     \biggl(-g^{ij}\,\partial_i\Phi\,\partial_j\Phi
           -{\mu^2\over\hbar^2}\,\Phi^2\biggr)&\cr
   &={1\over2}\int dt d^3x\, \sqrt{-g}\,
     \Phi\,\biggl(\qadratello
    \Phi-{\mu^2\over\hbar^2}\,\Phi\biggr)&(2.7)\cr
}
$$
where $g=\det g_{ij}$, $\qadratello$
is the Laplace--Beltrami operator in Kerr space--time,
$\mu/\hbar$ is the inverse of the Compton wavelength
of the scalar field and $g^{ij}$ the inverse Kerr metric.
At the basis of the adiabatic approximation is the hypothesis
that the heavy degrees of freedom ($\gamma$) appearing in
$S^{^M}$ vary slowly enough that the Hamiltonian can be treated
as constant, that is it is frozen at the value of $t$ at
which we want to evaluate it.
This means that in our scalar field Klein--Gordon equation
we just consider a static Kerr geometry in which case it
becomes separable [9]
and one may write:
$$
\Phi=\varphi(t)\,e^{i\,m\,\phi}\,S(\theta)\,R(r)
\ ,
\eqno(2.8)
$$
where the eigenfunctions for the angular part of the Kerr space
Klein--Gordon equation are the spheroidal harmonics and $R(r)$
will be the eigenfunctions solution to the radial equation which
can be obtained analytically [10] or in the WKB approximation [11],
depending on the relative masses of the scalar field and black
hole.
In our case we have $\mu M\ll\hbar$ and we may use the analytic
results previously obtained in [10],
in which case the  $S(\theta)$ become spherical harmonics and
for the radial part one obtains solutions with as boundary conditions
an outgoing wave at infinity and a downgoing wave at the event
horizon; such a solution corresponds to a particular mode of free
oscillation of a scalar field.
Let us note that for ``large'' distances from the event horizon
one obtains an analytic solution, resembling that for an electron
in the hydrogen atom, which can be expressed in terms of confluent
hypergeometric functions.
Similarly, near the event horizon one again obtains an analytic
solution which is expressible in terms of two other hypergeometric
functions, one of which is ingoing and the other is outgoing
at the event horizon.
There is a region, however, where the solution outgoing at infinity
can be matched to the near--the--horizon solutions.
If after the matching the coefficient of the solution outgoing
at the horizon vanishes, we have the desired solution corresponding
to the free oscillation of the scalar field.
Let us again emphasize that the above simplification for the
Klein--Gordon equation and the resulting relation to the
Schr\"odinger equation for the electron in the hydrogen atom
occurs when the mass of the scalar field (and its associated
energy) are much less than the inverse of the scalar field mass
($M/\hbar\ll 1/\mu$).
\par
Clearly, since one has an ingoing solution at the event horizon, one
must have a complex frequency $\omega=\sigma+i\beta$ ($\sigma$ and
$\beta$ real) and principal quantum number $\nu=n+l+1+\delta\nu$,
with $\delta\nu$ small and complex, $n$ integral and non--negative
(radial quantum number) and $l$ the angular momentum.
The relationship between the different quantities are given by [10]:
$$
{\mu^2\over\hbar^2}-\sigma^2={\mu^2\over\hbar^2}\,
\biggl({\mu\, M/\hbar\over n+l+1}\biggr)^2
\ ,
\eqno(2.9)
$$
so that $\sigma\simeq {\mu\over\hbar}$, and:
$$
i\,\beta={\delta\nu\over M}\,
\Biggl({\mu\, M/\hbar\over n+l+1}\Biggr)^3
\ ,
\eqno(2.10)
$$
with:
$$
\eqalignno{
 \beta= & {\mu\over\hbar}\,\Biggl({\mu\,M\over\hbar}\Biggr)^{4\,l+4}
          \Biggl({m\,a\over M}-2\,{\mu\over\hbar}\,r_+\Biggr)
          \,{2^{4\,l+2}(2\,l+1+n)!\over(l+1+n)^{2\,l+4}\,n!}\,
          \Biggl[{l!\over(2\,l)!\,(2\,l+1)!}\Biggr]^2\cr
        & \times\prod_{j=1}^l\Biggl[j^2\,\biggl(1-{a^2\over M^2}\biggr)
          +\biggl({m\,a\over M}-2\,{\mu\over\hbar}\,r_+\Biggr)^2\Biggr]
\ .     & (2.11)\cr}
$$
It is worth noting that the imaginary part is positive if:
$$
m>m_0\equiv2\,{\mu\, M\over\hbar}\,{r_+\over a}
\ ,
\eqno(2.12)
$$
which is the condition for having a classical super--radiant
mode, which is seen to irradiate from the black hole by a
Boyer--Lindquist observer even if it is locally purely
ingoing on the event horizon, and we also observe that $m_0$ is
very small in our approximation ($\mu M/\hbar\ll 1$).
It is this super--radiant mode which will dominate for sufficiently
low black hole masses.
Let us note that, in contrast with the hydrogen atom for which the
electron wave function must be regular at the origin, in our
case the inner boundary condition must correspond to radiation
down the black hole.
It is for this reason that the principal quantum number $\nu$ is
modified by the addition of $\delta\nu$ (small and complex);
clearly if the scalar field solution is prevented from approaching
the black hole $\delta\nu=0$ [10] with $n$, $l$, $m$ unchanged,
consistently with the adiabatic approximation,
and one just has the usual hydrogen-atom-like state.
\par
One may expand a generic scalar field $\Phi$ into the complete
set of spherical harmonics and radial solutions described above,
substitute such an expansion into Eq. (2.7) and perform the spatial
integration.
At this point we shall be left with the following expression:
$$
S^{^M}_{eff}={1\over 2}\int dt\,(\dot\varphi^2-\omega^2\,\varphi^2)
\ .
\eqno(2.13)
$$
Clearly one would have a different field $\varphi$ and frequency
$\omega_{n,l,m}(\gamma)$ ($\equiv\omega$) for each mode labelled
by $n,l,m$ numbers.
However we have just exhibited one, without specifying it further,
since we shall later (since $\mu M\ll\hbar$) just consider
two special modes:
the {\sl lowest} (and most) {\sl stable} mode and
the corresponding {\sl lowest} (and most) {\sl unstable} mode,
corresponding to $n=0$, $l=1$, $m=-1$ and $n=0$, $l=1$, $m=1$
respectively ($2p$ states of the hydrogen atom), which are
the ones of greatest interest and the generalization to an
arbitrary number of modes is straightforward.
\par
{}From the above one obtains a classical Hamiltonian $H^{^M}$:
$$
H^{^M}={1\over2}\,(\pi_\varphi^2+\omega^2\,\varphi^2_{})
\ ,
\eqno(2.14)
$$
where $\pi_\varphi$ is the momentum conjugate to $\varphi$, and,
through canonical quantization, a non--hermitian Hamiltonian
operator:
$$
\hat H^{^M}=-{\hbar^2\over2}\,{\partial^2\over\partial\varphi^2}
         +{\omega^2\over2}\,\varphi^2
\ .
\eqno(2.15)
$$
Now, as we mentioned in the Introduction, because of reparametrization
invariance the sum of the above matter Hamiltonian and the gravitational
one will be zero.
Nonetheless it can be shown [3,6] that for the matter--gravity
system in the semiclassical limit for gravity one obtains a
parametrization for the evolution of matter in which the latter follows
the former adiabatically.
One then obtains, neglecting fluctuations (adiabatic approximation),
a matter wave function
$\tilde\chi_s(\gamma,\varphi)$
which satisfies the usual equation of motion:
$$
\biggl[\hat H^{^M}-i\,\hbar\,{\partial\over\partial t}\biggr]\,
\tilde\chi_s=0
\ .
\eqno(2.16)
$$
As we shall see from the above, without going through the details
of the coupled
matter--gravity equations [3,6], one can extract both the induced
and dynamical phases.
\par
Indeed on defining a new matter wave function
$\chi(\gamma,\varphi)$
which only depends on $t$ through $\gamma^\lambda$:
$$
\eqalignno{
\tilde\chi_s & \equiv e^{-{i\over\hbar}\,\int^t dt^\prime\,
          \langle\,\hat H^{^M}\,\rangle
-i\,\int^t dt^\prime\,\dot\gamma^\lambda\,A_\lambda}
\,\chi                      \cr
             & \equiv e^{-{i\over\hbar}\,\int^t dt^\prime\,
                  \langle\,\hat H^{^M}\,\rangle}\,
                \tilde\chi \ \ , & (2.17)       \cr
}
$$
where the first exponential is the dynamical phase,
which is associated with the backreaction of matter
on gravity [12], and:
$$
\langle\,\hat H^{^M}\,\rangle=
{\langle\chi|\,\hat H^{^M}\,|\chi\rangle\over
\langle\chi|\chi\rangle}
\ ,
\eqno(2.18)
$$
with:
$$
\langle\chi|\chi\rangle=\int d\varphi\,\chi^\ast(\gamma,\varphi)
\,\chi(\gamma,\varphi)
\ ,
\eqno(2.19)
$$
the integral being over the different field modes.
The second exponent in Eq. (2.17) is the adiabatically induced phase
which can be determined by substituting for
$\tilde\chi_s$ in Eq. (2.16) obtaining:
$$
\biggl[\hat H^{^M}-\langle\chi|\hat H^{^M}|\chi\rangle-\hbar\,A_\lambda
-i\,\hbar\,\dot\gamma^\lambda\,{\partial\over\partial\gamma^\lambda}
\biggr]\,\chi=0
\ ,
\eqno(2.20)
$$
which after multiplying on the L.H.S. by $\chi^\ast$ and integrating
over the field modes leads to:
$$
A_\lambda=-i\,{\langle\chi|\,{\partial\over\partial\gamma^\lambda}\,
|\chi\rangle\over\langle\chi|\chi\rangle}
\equiv -i\,\langle\,{\partial\over\partial\gamma^\lambda}\,\rangle
\ .
\eqno(2.21)
$$
\par
Returning to $\hat H^{^M}$, we note that it, as a consequence of
the adiabatic approximation [13], is of the harmonic oscillator
form and we can introduce eigenstates $|N\rangle$ and eigenfunctions
$\chi_{_N}(\varphi)\equiv\langle\varphi|N\rangle$
($\tilde\chi_{_N}(\varphi)\equiv\langle\varphi|\tilde N\rangle$)
satisfying:
$$
\hat H^{^M}\,|N\rangle=E_{_N}\,|N\rangle
=\biggl(N+{1\over2}\biggr)\,\hbar\,\omega\,|N\rangle
\eqno(2.22)
$$
and on remembering that $\omega$ is complex one may introduce
a dual basis $\langle\bar N|$
and eigenfunctions:
$\bar\chi^\ast_{_N}(\varphi)\equiv\langle\bar N|\varphi\rangle$
satisfying:
$$
\langle\bar N|\,\hat H^{^M}=\langle\bar N|\,E_{_N}=
\langle\bar N|\,\biggl(N+{1\over2}\biggr)\,\hbar\,\omega
\ ,
\eqno(2.23)
$$
with:
$$
\langle\bar N|L\rangle=\delta_{N\,L}
\ ,
\eqno(2.24)
$$
together with a completeness relation:
$$
\sum_N|N\rangle\,\langle\bar N|=1
\ .
\eqno(2.25)
$$
\par
The above will allow us to obtain a different expression for
the adiabatic phase.
Indeed if for simplicity henceforth we take
$|\chi\rangle=|N\rangle$
we have:
$$
\eqalignno{
i\,(\partial_\lambda A_\mu-\partial_\mu A_\lambda) &
  = {\partial\over\partial\gamma^\lambda}\,\langle N|\,
    {\partial\over\partial\gamma^\mu}\,|N\rangle
   -{\partial\over\partial\gamma^\mu}\,\langle N|
    {\partial\over\partial\gamma^\lambda}\,|N\rangle       \cr
& = \langle\partial_\lambda N|\partial_\mu N\rangle
   -\langle\partial_\mu N|\partial_\lambda N\rangle      \cr
& =\sum_{L\not= N}[
\langle\partial_\lambda N|L\rangle\,\langle\bar L|\partial_\mu N\rangle
-\langle\partial_\mu N|L\rangle\,\langle\bar L|\partial_\lambda N\rangle]
\cr
& \approx  \sum_{L\not= N}[
   \langle\partial_\lambda N|L\rangle\,\langle L|\partial_\mu N\rangle
   -\langle\partial_\mu N|L\rangle\,\langle L|\partial_\lambda N\rangle]
   \cr
& \approx \sum_{L\not= N}{1\over\mu^2\,(N-L)^2}\,\biggl[
    \langle N|\,\biggl({\partial\hat H^{^M}\over\partial\gamma^\lambda}
    \biggr)^\dagger\,|L\rangle\,\langle L|\,
    {\partial\hat H^{^M}\over\partial\gamma^\mu}\,|N\rangle    \cr
&\ \ \ \ \ \ \ \ \ \ \ \ \
   - \langle N|\,\biggl({\partial\hat H^{^M}\over\partial\gamma^\mu}
     \biggr)^\dagger\,|L\rangle\,\langle L|\,
     {\partial\hat H^{^M}\over\partial\gamma^\lambda}\,|N\rangle
\ ,
&(2.26)     \cr}
$$
where we have used the fact that $\beta$, Eq. (2.11), is small and we
observe that, since the matter Hamiltonian is complex (presence of a
gravitational horizon), the matter wave function cannot be made real
and therefore the above is non--zero.
Then, if we consider a suitable closed cycle ${\cal C}$ in the parameter
space of the Kerr black hole metric, on using
Eq. (2.26) we obtain for the non--dynamical phase factor in Eq. (2.17):
$$
\eqalignno{
\oint_{\cal C}d\gamma^\mu\,\langle N|\,{\partial\over\partial\gamma^\mu}
     \,|N\rangle\approx
- & \sum_{L\not= N}
  \int\int_\Sigma{da\,dM\over\mu^2\,(N-L)^2}\,
    \langle\tilde N|\,\varphi^2\,|\tilde L\rangle
    \langle\tilde L|\,\varphi^2\,|\tilde N\rangle        \cr
& \times
    \biggl[\biggl(\omega\,{\partial\omega\over\partial a}\biggr)^\ast
    \,\omega\,{\partial\omega\over\partial M}
    -\biggl(\omega\,{\partial\omega\over\partial M}\biggr)^\ast
    \,\omega\,{\partial\omega\over\partial a}\biggr]
    \ ,
& (2.27)\cr}
$$
where $\Sigma$ is the 2-surface enclosed by ${\cal C}$ in the
$(a,M)$ plane.
Concerning such a contour ${\cal C}$ it can be put into correspondence
with a closed path ${\cal C}^\prime$ in the $\Omega$
(angular velocity) and $J$ (angular momentum) plane.
Since one can relate ($a,M$) to ($\Omega,J$)
[7], such a contour (smooth closed curve) representing a reversible
cycle may be replaced (as in the usual thermodynamics) by a
reversible zigzag closed path consisting of alternate thermodynamic
adiabatic and isothermal portions performed by mechanisms
such as those described elsewhere [7].
\par
Although it is not our scope to discuss the interaction
of our scalar field with such mechanisms we feel a few
comments are in order.
It is clear that they could lead to a modification of the
boundary conditions.
For example during the thermodynamic adiabatic process
(constant entropy for the black hole)
the scalar field will not leak in the black hole
($\delta\nu=0$) and one will just have a stable bound state
(as an electron in a hydrogen atom).
Thus for certain portions of the cycle $\beta$ could be zero,
therefore in our expressions one could just replace it by some
sort of averaged (over successive adiabatic and isothermal portions)
non zero value.
However since we are only interested in order of magnitude
estimates we shall not pursue this further.
Lastly let us address the issue of thermodynamic stability
and the possibility of quasistatic processes.
Specific heats of relevance for the above cycles are the
specific heat at constant $\Omega$ and that at constant $J$.
The former is always negative, however a way of avoiding
this difficulty by limiting the energy of the
``heat bath'' has already been pointed out [7];
concerning the latter specific heat, it can be rendered positive
by taking $a/M$ greater than a minimum value of the order
of 0.7 [7].
\par
We may now proceed to explicitly apply the above results to a
particular case.
Regarding the choice of the matter wave function we observe
the quantization of $\varphi$, Eq. (2.15),
is in reality a second quantization.
Thus if we consider
a superposition of eigenmodes $\tilde\chi_{_N}$
which is associated with a coherent state, in the limit for
$\hbar\to 0$, such a state will lead to the Hamilton--Jacobi
equation of motion for the classical field $\varphi$.
Such a coherent superposition of harmonic oscillator energy
eigenfunctions $\tilde\chi_{_N}$ is given by:
$$
\tilde\chi_c=\biggl({\sigma\over\hbar\,\pi}\biggr)^{1/4}
             \sum_{L=0}^\infty{1\over(2^L\,L!)^{1/2}}\,
             \biggl({\omega\over\hbar}\,\varphi_0\biggr)^{L/2}
             e^{-{\omega\over4\,\hbar}\,\varphi_0^2}
             e^{-i\,(L+1/2)\,\omega\, t}\,\tilde\chi_{_L}
\ ,
\eqno(2.28)
$$
which, for $t=0$, corresponds to a minimum wavepacket centered on
$\varphi_0$:
$$
\tilde\chi_0=\biggl({\sigma\over\hbar\,\pi}\biggr)^{1/4}
             e^{-{\omega\over2\,\hbar}\,(\varphi-\varphi_0)^2}
       \equiv\langle\varphi|0\rangle
\ .
\eqno(2.29)
$$
For $t\not= 0$, since $\omega$ is complex, one will obtain a
wavepacket oscillating with the classical frequency $\sigma$
(the metric being essentially classical) about $\varphi=0$
and with amplitude $e^{\beta t}\varphi_0$.
This essentially occurs because the coefficients in Eq. (2.28) have
a maximum at $L=N_c={\omega\over2\hbar}\varphi_0^2e^{2\beta t}$
and their squares obeys a Poisson distribution with average
value $N_c$ and standard deviation $\sqrt{N_c}$.
Thus, if $\beta$ is negative ({\sl stable} mode) the relevant value
$L$ ($t\gg 1/\beta$) will be $L=0$ and the state  $\chi_c$
reduces to a minimum wavepacket,
whereas for $\beta$ positive ({\sl unstable} mode) $N_c$ will
increase exponentially ($N_c\sim e^{2\beta t}$), corresponding,
since $\langle\hat H^{^M}\rangle\simeq\hbar\omega N_c$,
to the creation of additional scalar field quanta.
The existence of such an exponential increase for a mode satisfying the
super--radiance condition Eq. (2.12) is due the rotational degree of
freedom of the black hole (in fact, in the limit $a\to 0$
the condition Eq. (2.12) can never be satisfied and so $\beta$ is
always negative).
For suitable values of $M$ the growth time of the instability
is much less than the evaporation time scale in agreement with
the adiabatic approximation.
\par
We first consider the case of a minimum wavepacket
for the state $n=0$, $l=1$, $m=-1$ in $\sigma$ and $\beta$
(lowest stable mode) and keep terms to lowest order in $\mu M$
and $\hbar$ obtaining from Eq. (2.27) with $N=0$:
$$
\oint_{\cal C} d\gamma^\mu\,
\langle0|\,{\partial\over\partial\gamma^\mu}\,|0\rangle
\approx -i\,{{\cal A}\over384}\,{\mu^2\over\hbar^2}\,
\biggl({\mu\, M\over\hbar}\biggr)^8
\ ,
\eqno(2.30)
$$
where ${\cal A}$ is the area in the $(a,M)$ plane enclosed by the
cycle ${\cal C}$.
One may now consider the case of the lowest unstable mode:
$n=0$, $l=1$, $m=1$.
For such a case all the diverse energy eigenfunctions in
Eq. (2.28) will contribute.
However, for our purpose of obtaining an estimate, it will be
sufficient just to consider the term in the sum (2.28) whose
coefficient is the maximum, that is $L=N_c$.
Thus we simply set $N=N_c$ in Eq. (2.27) obtaining:
$$
\oint_{\cal C} d\gamma^\mu\,
\langle N_c|\,{\partial\over\partial\gamma^\mu}\,|N_c\rangle
\approx -i\,{{\cal A}\over384}\,{\mu^2\over\hbar^2}\,
\biggl({\mu\, M\over\hbar}\biggr)^8\, N_c^2
\ .
\eqno(2.31)
$$
\par
The presence of such contributions  will modify the
energy levels $E_{_N}$, Eq. (2.22), for the matter Hamiltonian
since we solved for them in the adiabatic approximation.
The effect of the inclusion of the adiabatically induced
phase is easiest seen by writing the corresponding
semiclassical quantization condition for the matter
wave function:
$$
i\,\oint_{\cal C}\pi_\varphi\, d\varphi +\hbar \,
\oint_{\cal C} d\gamma^\mu\,
\langle N|\,{\partial\over\partial\gamma^\mu}\,|N\rangle
=2\,\pi\, i\,\biggl(N+{1\over2}\biggr)\,\hbar
\ ,
\eqno(2.32)
$$
which in the absence of the second term in the L.H.S.
leads to the usual energy spectrum $E_{_N}$.
\par
We note that the black hole wave function (which, as we mentioned
in the Introduction, we need not know beyond observing that
in the semiclassical limit it is related to the exponential
of the classical gravitational action, that is the area of the horizon)
will acquire an equal
and opposite phase to that of matter [3,4,6] and it is further
worth observing that during a cycle in parameter space
the net work $W$ extracted from the black hole is given by [7]:
$$
W=-\oint\Omega\, dJ
\ ,
\eqno(2.33)
$$
where $\Omega$ is the angular velocity which is related to the
surface area $A$ and entropy $S$ of the black hole through:
$$
\Omega={4\,\pi\, a\over A}={\kappa_{_B}\,\pi\, a\over\hbar\, S}
\ ,
\eqno(2.34)
$$
where $\kappa_{_B}$ is the Boltzmann constant.
Since $a$ and $\Omega$ return to their initial values at the
end of the cycle ${\cal C}$, from Eq. (2.34) there in no net change
in the classical entropy of the black hole,
as is expected for any reversible thermodynamic process.
However as we have mentioned, the wave function of the black hole
acquires an additional phase during the cycle which reflects the
anholonomy of the process and in particular this arises because of
the instability of the scalar field bound to the black hole.
This is precisely due to the fact that we have imposed the
boundary condition that on the event horizon the scalar field
is purely infalling, that is it leaks into the black hole,
thus giving rise to an irreversibility and an increase in the
{\sl quantum} black hole entropy (Berry phase) even if at the
end of the cycle we have returned to the same values of $(a,M)$.
That is the black hole action is no longer just the
geometric surface term associated with the value of $(a,M)$ but one
has an additional contribution due to the adiabatically
induced phase, it thus appears that the black hole seems
to retain a ``memory'' of the cycle.
It is interesting to speculate that in general irreversible
processes are related to either quantum mechanical instabilities
or the generation of anholonomies due to topological defects
or non trivial topological structures which prevent the shrinking
of the closed cycle to a point (for example, the generation of
defects in a crystal caused by the cyclic variation of macroscopic
variables).
\par
One last point one should address concerns the validity of the
adiabatic approximation which requires that the adiabatically induced
phase be smaller than the dynamical phase.
This means that the time dependence induced by the variations of
$M$ and $a$ be smaller than that due to $\hat H^{^M}$.
In particular the former is given by
$-i\hbar\dot\gamma^\mu{\partial\tilde\chi\over\partial\gamma^\mu}$
which can be estimated as:
$$
\eqalignno{
& \hbar\,\biggl[\dot\gamma^\mu\,\dot\gamma^\lambda\,
    \sum_{L\not= N}{1\over\mu^2\,(N-L)^2}\,\langle N|\,
    \biggl({\partial\hat H^{^M}\over\partial\gamma^\lambda}\biggr)^\dagger
    \,|L\rangle\,\langle L|\,
    {\partial\hat H^{^M}\over\partial\gamma^\mu}\,|N\rangle
    \biggr]^{1/2}                               \cr
&  =\,\hbar\,\biggl[\dot\gamma^\mu\,\dot\gamma^\lambda \,
    \sum_{L\not= N}{1\over\mu^2\,(N-L)^2}
    \langle N|\,\omega^\ast\,{\partial\omega^\ast\over\partial\gamma^\mu}
    \,\varphi^2\,|L\rangle\,\langle L|
    \,\omega\,{\partial\omega\over\partial\gamma^\lambda}
    \,\varphi^2\,|N\rangle\biggr]^{1/2}
\ ,
&(2.35)\cr}
$$
which for the case $N=0$ leads to, taking $\delta\dot M\sim\delta\dot a$:
$$
\hbar\,\biggl[\dot\gamma^\mu\,\dot\gamma^\lambda
     \,{1\over4\hbar}\,{\partial\omega^\ast\over\partial\gamma^\mu}
     \,{\partial\omega\over\partial\gamma^\lambda}
     \,\langle0|\,\varphi^2\,|2\rangle\,\langle2|\,\varphi^2\,|0\rangle
     \biggr]^{1/2}
\approx\delta\dot M\,{\mu\over8\,\sqrt{2}}\,
\biggl({\mu\, M\over\hbar}\biggr)
\ .
\eqno(2.36)
$$
Such a term compared with the dynamical phase
($\langle\hat H^{^M}\rangle\sim\mu$) is very small for $\mu M\ll\hbar$.
This is just a statement of the adiabatic approximation.
\par
For the case of the lowest unstable mode we may proceed as before
taking $N=N_c$, and obtain:
$$
\eqalignno{
\hbar\, & \biggl[\dot\gamma^\mu\,\dot\gamma^\lambda
     \,{1\over4\hbar}\,{\partial\omega^\ast\over\partial\gamma^\mu}
     \,{\partial\omega\over\partial\gamma^\lambda}
\,\biggl(\bigl|\langle\tilde N_c|\,\varphi^2\,|\tilde N_c-2\rangle\bigr|^2
+ \bigl|\langle\tilde N_c|\,\varphi^2\,|\tilde N_c+2\rangle\bigr|^2\biggr)
     \biggr]^{1/2}                           \cr
& \approx\delta\dot M\,{\mu\over8\,\sqrt{2}}\,
\biggl({\mu\, M\over\hbar}\biggr)
  \, N_c
\ ,&(2.37)
\cr}
$$
which is again small since it must be compared with
$\langle\hat H^{^M}\rangle\sim\mu N_c$.
\bigskip
\centerline{\bf III. Discussion and Conclusions}
\medskip
In the previous section we have considered a scalar field bound
to a rotating black hole which undergoes a cycle in the
$(a,M)$ plane corresponding to a cycle in the $(\Omega,J)$ plane.
In order to obtain some understanding of the quantities involved
it is useful to obtain some order of magnitude estimates.
\par
In order to perform the requested process quasi--statically
(or almost adiabatically, in the quantum mechanical sense)
one must always have a quasi--equilibrium between the black hole
and the heat bath by means of which the cycle is performed.
This implies that the characteristic time for the changes
during the cycle must be much longer than the thermal relaxation
time of the system [7].
This would mean that, since the evaporation time scale is [14]:
$$
\tau_{evap}\approx
(10^{17}\, {\rm sec})\,\biggl({M\over2\cdot10^{15}\,{\rm g}}\biggr)^3
\equiv(3\cdot10^{27}\,{\rm cm})\,\biggl({M\over M_u}\biggr)^3
\ ,
\eqno(3.1)
$$
where $M_u$ ($=2\cdot10^{15}$ g) is the mass of a black hole with
the same lifetime as the universe,
one must have a black hole mass of the order
of $10^{4}\,{\rm g}$ for the evaporation time to be of the order of
the $\pi^0$ lifetime ($\sim 10^{-16}\,{\rm sec}$).
Let us note however that it may well be that if the number of species
of particles is very large already for $M\sim10^{14}\,{\rm g}$,
the relevant time scale can be of the order of the strong interaction
one ($10^{-23}\,{\rm sec}$), thus the above Eq. (3.1) would not be
applicable.
Let us nonetheless use it while bearing the above in mind especially
since for $M\sim10^4\,{\rm g}$ one would have ($\mu_\pi=$ pion mass)
$\mu_\pi\, M\sim10^{-77}\,{\rm cm}^2$ which is $\ll\hbar$
($=2.6\cdot10^{-66}\,{\rm cm}^2$) as required.
\par
We observe that in our approach the scalar field is localized near
the horizon indeed:
$$
r_+\sim(4\cdot10^{-13}\,{\rm cm})\,\biggl({M\over M_u}\biggr)
\ ,
\eqno(3.2)
$$
while the Bohr radius $a_0$ for the scalar field bound to the black
hole is:
$$
a_0\sim{4\,\pi\,\hbar^2\over\mu^2\, M}\approx(10^{-12}\,{\rm cm})
\,\biggl({M_u\over M}\biggr)\,\biggl({\mu_\pi\over\mu}\biggr)^2
\ ,
\eqno(3.3)
$$
which can be comparable.
This of course means that if one wanted to study the backreaction
of matter on the gravitational field [12] it would be sufficient to
examine the Einstein equations where the scalar field has support,
that is near the horizon.
\par
The results we have obtained are quite different depending
on whether the scalar field bound states are stable or not,
that is $\beta$ negative or positive respectively.
The fastest growing instability corresponds to $l=1$, $m=1$, $n=0$
in which case the growth time $\tau$ is given by:
$$
\eqalignno{
\tau\sim{1\over\beta} & =
24\,{\hbar\over\mu}\,\biggl({M\over a}\biggr)
\,\biggl({\mu\, M\over\hbar}\biggr)^{-8} \cr
& \approx(6\cdot10^{-12}\,{\rm cm})\,\biggl({M\over a}\biggr)
  \,\biggl({\mu_\pi\over\mu}\biggr)
  \,\biggl({\mu_\pi\, M_u\over\mu\, M}\biggr)^8                      \cr
& \approx(2\cdot10^{-22}\,{\rm sec})\,\biggl({M\over a}\biggr)
  \,\biggl({\mu_\pi\over\mu}\biggr)
  \,\biggl({\mu_\pi\, M_u\over\mu\, M}\biggr)^8
\ ,
&(3.4)
\cr}
$$
which for $M\sim10^4\,{\rm g}$, $\mu=\mu_\pi$ and $a/M\sim 0.8$
leads to $\tau\sim5\cdot10^{69}\,{\rm sec}$, which is much longer than
the lifetime of the universe, that is $\beta$ is very small as
expected.
However for more massive black holes the value changes rapidly because
of the power $8$ in the Eq. (3.4).
\par
We have noted that the adiabatically induced phase by a cyclic
variation of $a$ and $M$ leads to an energy shift of the scalar
field bound states.
Indeed from Eq. (2.32) we have:
$$
2\,\pi\, i\, {E_{_N}\over\sigma}=
2\,\pi\, i\,\biggl(N+{1\over2}\biggr)\,\hbar
-\hbar\,\oint_{\cal C}d\gamma^\mu\,\langle N|\,
{\partial\over\partial\gamma^\mu}\,|N\rangle
\ ,
\eqno(3.5)
$$
thus obtaining for the stable case ($N=0$):
$$
\eqalignno{
E_0  & ={1\over2}\,\hbar\,\sigma+{\hbar\,\sigma\over2\,\pi}
 \,{{\cal A}\,\mu^2\over384\,\hbar^2}
 \,\biggl({\mu\, M\over\hbar}\biggr)^8
={1\over2}\,\hbar\,\sigma\,\biggl[1+10^{-3}\,{{\cal A}\over M^2}
 \,\biggl({\mu\, M\over\mu_\pi\, M_u}\biggr)^{10}\biggr] \cr
& \equiv {1\over2}\,\hbar\,\sigma[1+\delta]
\ ,
&(3.6)  \cr}
$$
which is in general a small ($<10^{-3}$) correction.
For the unstable case one will obtain:
$$
E_{_N}={1\over2}\,N_c\,\hbar\,\sigma\,\biggl[2+10^{-3}
    \,{{\cal A}\over M^2}
    \,\biggl({\mu\, M\over\mu_\pi\, M_u}\biggr)^{10}\,N_c\biggr]
\ ,
\eqno(3.7)
$$
which can lead, for suitable values of the parameters and $N_c$ large,
to a significant shift.
\par
Concerning the validity of the adiabatic approximation which is
obtained by comparing the time dependence due to the variation
of the metric with $\langle\hat H^{^M}\rangle$, one has for the stable
case:
$$
{\hbar\,\biggl[\dot\gamma^\mu\,\dot\gamma^\lambda
\,{1\over\hbar}\,{\partial\omega^\ast\over\partial\gamma^\mu}
\,{\partial\omega\over\partial\gamma^\lambda}
\,\langle0|\,\varphi^2\,|2\rangle\,\langle2|\,\varphi^2\,|0\rangle
     \biggr]^{1/2}
\over\langle\,\hat H^{^M}\,\rangle}
\approx{\delta\dot M\over8\,\sqrt{2}}\,{\mu\, M\over\hbar}
={\delta\dot M\over10}\,\biggl({\mu\, M\over\mu_\pi\, M_u}\biggr)
\ ,
\eqno(3.8)
$$
which in general is very small as can be verified on estimating
$\delta\dot M$ as :
$$
\delta\dot M\sim\biggl({\partial\tau_{evap}\over\partial M}\biggr)^{-1}
\approx 3\cdot10^{-41}\,\biggl({M_u\over M}\biggr)^2
\ .
\eqno(3.9)
$$
The conclusions are unchanged for the unstable case.
This of course confirms the choice of the scalar field being the
``fast'' variable and the black hole parameters the ``slow'' ones.
\par
Let us then summarize: we considered a scalar field
bound to a Kerr black hole whose geometry is varying through
some external mechanism,
that is we considered an ``open'' system.
We then identified, for the wave function obtained after second
quantization of the scalar field, the adiabatically and dynamically
induced phases.
Since for a scalar field bound to a Kerr hole one has both stable
and unstable modes we examined two cases.
One associated with the stable solutions in which there is no increase
in the number of scalar particles orbiting the black hole and
the other associated with instability in which the number of particles
increases exponentially.
Clearly the latter is the more interesting of the two.
Indeed, for a closed cycle we noted that the quantization
condition is modified due to the contribution of the adiabatically
induced phase and for the case of the unstable levels the effect
is more pronounced.
Concerning the black hole wavefunction we noted that it acquires
an equal and opposite phase to that of matter during the cycle,
which reflects the anholonomy of the process, because of the
instability of the scalar field.
This phase can be considered as an increase
in the black hole entropy (action).
It is interesting to speculate that in general irreversible processes
are related either to quantum mechanical instabilities or the
generation of anholonomies due to topological defects or non trivial
topological structures.

\bigskip
\centerline{\bf References}
\medskip

\item{[1]} B. S. DeWitt, Phys. Rev. {\bf 160}, 1113 (1967). \par

\item{[2]} R. Brout, Found. Phys. {\bf 17}, 603 (1987);\hfill
           \break
           Z. Phys. B {\bf 60}, 359 (1987).               \par

\item{[3]} R. Brout and G. Venturi,  Phys. Rev. D {\bf 31},
           2436 (1989).                                     \par

\item{[4]} C. A. Mead and D. G. Truhlar, J. Chem. Phys.
           {\bf 70}, 2284 (1974);\hfill
           \break
           C. A. Mead, Chem. Phys. {\bf 49}, 23, 33 (1980). \par

\item{[5]} M. V. Berry, Proc. R. Soc. London A{\bf 392}, 45
           (1984). \par

\item{[6]} For a general treatment see G. Venturi in {\it Differential
           Geometric Methods in Theoretical Physics}, edited by L.-L.
           Chau and W. Nahm, Plenum Press, 1990.\par

\item{[7]} I. Okamoto and O. Kaburaki, Mon. Not. R. astr. Soc.
           {\bf 247}, 244 (1990);\hfill\break
           O. Kaburaki and I. Okamoto,  Phys. Rev. D {\bf 43},
           340 (1991).   \par

\item{[8]} R. H. Boyer and R. W. Lindquist, J. Math. Phys.
           {\bf 8}, 265 (1967).  \par

\item{[9]} D. R. Brill, P. L. Chrzanowski, C. M. Pereira,
           E. D. Fackerell and J. R. Ipser, Phys. Rev. D {\bf 5},
           1913 (1972). \par

\item{[10]} S. Detweiler, Phys. Rev. D {\bf 22}, 2323 (1980). \par

\item{[11]} T. J. M. Zouros and D. M. Eardley, Ann. Phys. {\bf 118},
            139 (1979).  \par

\item{[12]} By this we mean that the Born--Oppenheimer (or adiabatic)
            approximation to Einstein equations leads to gravity
            being driven by the (quantum) mean energy--momentum
            tensor of matter (see ref. [2], [3] and [6]).
            This is a generalization of the work of T. Banks
            (Nucl. Phys. {\bf B249}, 332 (1985)) which did not
            include the effect of the feedback of matter on
            gravity.\par

\item{[13]} By this we mean that in the scalar field Hamiltonian
            the gravitational degrees of freedom ($\gamma$) are
            frozen at the value of $t$ at which we want to evaluate
            it.\par

\item{[14]} S. W. Hawking, Commun. Math. Phys. {\bf 43}, 199 (1975).
            \par

\end